# Plastic or Viscous? A Reappraisal of Yielding in Soft Matter

Yogesh M. Joshi[1*] and Alexander Ya. Malkin[2*]

1 Department of Chemical Engineering, Indian Institute of Technology Kanpur, Kanpur, Uttar Pradesh 208016, India

2 A. V. Topchiev Institute of Petrochemical Synthesis, Russian Academy of Science

Russia, 119991, Moscow, 29. Leninskii prospect

**Abstract:**

Many soft jammed materials, such as pastes, gels, concentrated emulsions, and suspensions, possess a threshold stress, known as yield stress, that must be exceeded to cause permanent deformation or flow. In rheology, the term plastic flow is commonly used to describe continuous flow (unbounded increase in strain with time) that a material undergoes above a yield stress threshold. However, in solid mechanics, plasticity refers to irreversible but finite, rate-independent deformation (strain that does not evolve with time). In addition, many soft materials exhibit viscosity bifurcation, a prominent thixotropic signature, which further complicates the definition and interpretation of yield stress. The threshold stress at which viscosity bifurcation occurs is also termed a yield stress, even though deformation below this threshold is not purely elastic, while above this threshold, the material flows homogeneously with a constant shear rate. This paper revisits these critical issues by analyzing the rheological and solid mechanics perspectives on plasticity. The insights presented here are intended to address certain terminological ambiguities for interpreting flow in soft jammed materials.

*Corresponding authors**;** joshi@iitk.ac.in and alex_malkin@mig.phys.msu.ru

## 1. Introduction

How materials deform under stress, whether elastically, plastically, or they undergo continuous flow, has long been a subject of scientific interest. The nature of deformation has significant consequences across fields ranging from metallurgy to soft matter while designing materials for various applications. Bingham,[1] in his pioneering publication, introduced the term "plastic flow" into rheology in relation to media capable of flow. He unambiguously demonstrated that while squeezing an aqueous clay suspension through a capillary, there exists a critical stress threshold beyond which volumetric flow occurs. He reported that below that threshold, flow was not possible. This threshold – the "yield point" $\sigma_y$ – separates the region of the solid-like state (at stresses below $\sigma_y$) from the region in which flow occurs when stress exceeds $\sigma_y$. His experiments with concentrated pitch suspensions provided clear evidence of this behavior. Subsequently, Bingham[2] distinguished the difference between the concepts of plasticity and fluidity. Bingham's description of visco-plastic fluids was widely applied to explain the rheological behavior of a variety of multicomponent systems, such as polymeric, colloidal and non-colloidal suspensions, as well as emulsions exhibiting similar characteristics. Meanwhile, around the same time, the concept of the plasticity of solids was developed in the works of HMH (Huber,[3] von Mises,[4] and Hencky[5]), culminating in a rigorous theory by Nadai.[6] These works led to a criterion for the transition from elastic to plastic behavior in solids. This criterion, termed as the HMH criterion, states that yielding occurs when the accumulation of maximum distortion strain energy exceeds a critical value. In this sense, the criterion both identifies the physical cause of the elastic-to-plastic yielding transition and specifies it to be the yielding condition. Interestingly, the term "plasticity" is being used interchangeably across rheology and solid mechanics communities, despite their distinct origins, which often leads to confusion. In rheology, it typically refers to irreversible, continuous, and homogeneous flow beyond a yield stress. On the other hand, in solid mechanics, it denotes irreversible but finite, rate-independent deformation, which may not be homogeneous. This conceptual overlap, despite being independently well-established in their respective domains, confounds critical physical distinctions between the two phenomena: finite plastic deformation in solids versus continuous flow in fluids. The matter becomes critical when soft materials also show plastic behavior as defined in solid mechanics. Considering this, the purpose of this perspective is to clarify these distinctions and highlight where the definitions align, where they diverge, and how they should be interpreted in the context of soft materials. In this perspective, references to plasticity in metals are employed selectively as a comparative framework to sharpen the distinctions, although the central emphasis of the discussion remains on yielding in soft materials.

To understand the origin of this debate, it is important to understand how the concepts of plastic flow and plastic deformation evolved independently. These observations refer to two fundamentally different dynamics, which is commonly termed as plasticity. The first describes a solid-to-liquid transition; the second, on the other hand, pertains to a solid-to-plastic transition, which is actually a solid-to-solid transition. Therefore, strictly speaking, in the first case, the use of the term plasticity may not be accurate, and it is more appropriate to describe such materials as yielding fluids or yield stress liquids, i.e., solids that transition to a fluid state beyond a threshold stress, rather than visco-plastic materials.[7]

It is also essential to clarify the distinction between elasticity, plasticity, and viscosity. Under constant applied stress, elastic as well as plastic materials undergo an instantaneous deformation (excluding possible viscoelastic effects), which depends on the magnitude of the stress but, in principle, remains unchanged over time. When stress is removed, deformation in elastic materials is completely recoverable, while that in plastic materials is not. On the other hand, deformation in viscous materials increases continuously with time, and the corresponding rate of strain is influenced by the applied stress. Interestingly, in both cases, the point of transition from the solid state is still governed by the same criterion, the HMH criterion, originally proposed for solids, which is applicable in multi-dimensional deformations of yielding fluids.[8-9]

If the material possessing yield stress has thermal constituents, and if the state of a material is out of thermodynamic equilibrium, the material tends to be thixotropic. Such a thixotropic material, once yielded, can reform its structure over time when the applied stress reduces below a certain threshold. This phenomenon is typically known as viscosity bifurcation,[10-11] and the corresponding threshold stress has also been termed as yield stress. However, the yield stress associated with viscosity bifurcation does not mark a transition from an elastic solid, as defined by the HMH criterion. Instead, it represents competition between structural recovery and its breakdown for a given magnitude of deformation field.[12] Such recovery of structure can make it difficult to determine whether the yielding in thixotropic materials represents a transition similar to that observed in classical plasticity, as it involves a transition either to diminishing shear rate or constant shear rate, respectively, above, below, and above the critical stress. The purpose of this paper is to discuss what happens to a material when it is subjected to stress exceeding the yield stress. We especially distinguish between continuous flow and true plasticity phenomena. By clearly delineating between plasticity in solids and flow in yield stress fluids, this article aims to address an issue of terminology used in soft materials that possess yield stress.

## 2. Yielding in Soft Materials

Yielding in soft materials originates from the presence of an internal structure that allows the material to resist irreversible or viscous deformation up to a critical stress threshold. Below this threshold, the material behaves like an elastic solid [13]. Above the critical stress, the structure breaks down, causing flow. Such behavior is commonly observed in many soft matter systems, such as gels, pastes, emulsions, suspensions, etc.[14] The structural arrest responsible for this behavior arises from a combination of intermolecular interactions, including chemical bonding, hydrogen bonding (as seen in protein-polysaccharide complexes,[15] electrostatic attractions, van der Waals forces, steric hindrance, depletion interactions, and hydrophilic or hydrophobic interactions.[16-18] The specific rheological properties of materials with arrested state depend on the nature and strength of these interactions, especially the ability to exhibit a solid-like response at low stress. The mechanical strength of the arrested state also depends on particle–solvent and particle–particle interactions, which define the robustness of the gel or network structure.[19-22] An important rheological characteristic of such structurally arrested soft materials is their solid-like response over a long time (or at low frequencies), often indicated by a frequency-independent elastic modulus.[7, 23] However, there persists a long-standing debate whether this solid-like behavior truly represents a permanent arrested state (with true elastic behavior) or the material undergoes ultra-slow flow below the yield stress threshold. This raises the fundamental question: does a true yield stress exist, or is flow merely delayed at very low stresses and long timescales?[24-25] The answer depends critically on the nature of the material's constituents. Several topical reviews have addressed the evolving understanding of the transition from a solid to a fluid state in yielding materials.[13-14, 26-34]

Many materials that possess yield stress could also be out of thermodynamic equilibrium. If the constituents of such materials are thermal, their microstructure evolves to lower the Gibbs free energy, and hence they show inherent time dependence.[12] The mobility of the constituents of the material determines the microstructural evolution. These materials are inherently thixotropic.[17, 35-36] Under applied stress (or applied deformation field), their structure undergoes progressive breakdown (a phenomenon known as rejuvenation), causing a time-dependent decrease in viscosity, inducing flow. However, the structure can rebuild (a phenomenon termed as physical aging) over time when the stress is removed or reduced. This time and stress-dependent competition between aging and rejuvenation plays a critical role in the yielding behavior of such systems. This fundamental distinction has significant implications for yielding behavior.[10, 37] If the constituents of a yield stress fluid are athermal, lacking Brownian motion, then, despite the system being out of thermodynamic equilibrium, microstructural evolution cannot occur. In such systems, the yielding transition is governed solely by direct mechanical interactions. Such materials show elastic solid-like response until stress exceeds a critical value. Such behavior is common in granular or jammed athermal suspensions. Consequently, these materials are non-thixotropic.

In materials with athermal constituents (non-thixotropic materials), yielding tends to be abrupt and immediately recoverable upon unloading. In thixotropic materials, progressive decreases in stress to vanishing values can cause viscous or irreversible flow [7, 20]. Such behavior is a departure from Bingham's classical yield-stress concept, which requires the yield point to be treated as a fixed material property. The two competing processes, namely the breakdown of the structured state of a material and the formation of structure for a given strength of deformation field govern evolution of relaxation times and the frequency-dependent behavior of the elastic modulus.[38] Figure 1 illustrates this dynamic nature of structure formation, highlighting the evolution frequency dependence of elastic modulus over time.

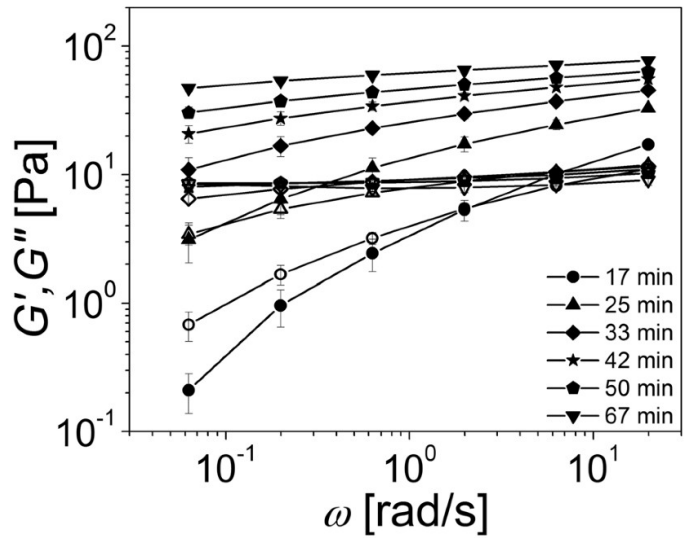

**Figure 1:** Evolution of storage (filled symbols) and loss modulus (open symbols) for a clay suspension at different aging times elapsed since shear melting. It can be seen that both the moduli increase with aging time, and its dependence on frequency becomes weaker. Increase in storage moduli suggests microstructural build-up, and its weakening dependence on frequency suggests a progressively strengthening structurally arrested state. Reproduced with permission from Jatav and Joshi.[39] Copyright 2014 Society of Rheology.

Thixotropic yield stress fluids may exhibit viscosity bifurcation, wherein the material shows a continuous increase in viscosity (owing to the evolution of structure) when applied stress is below a certain threshold.[10-11, 35, 40-43] A continuous increase in viscosity under constant stress causes a progressive decrease in shear rate over time. Ultimately, a state is reached where the viscosity becomes so high that any further deformation ceases. When material is subjected to stresses exceeding this threshold, the material continues to flow, eventually reaching a steady shear rate. During this process, viscosity changes continuously until it attains a constant value.[44] This behavior is illustrated in Fig. 2. In the literature, the corresponding threshold stress has also been termed as yield stress,[10] although the deformation that the material undergoes when applied stress is below the threshold is not entirely elastic.[10, 12, 35] Furthermore, after shear melting, if the thixotropic yield stress material is kept under rest conditions, the threshold value of stress below which the material achieves a steady state may continue to increase, leading to a time-dependent increase in yield stress as shown in Fig. 3.[10, 12, 35, 41-42] The soft glassy materials, which are considered to be inherently thixotropic,[17] are known to undergo evolution of relaxation time as a function of their age. This functional dependence can be obtained by performing creep experiments at different ages, followed by time-aging time superposition.[45-46] Typically, it has been observed that with an increase in creep stress, the dependence of relaxation time on age weakens and eventually ceases at a certain critical stress.[45, 47] This critical stress has also been termed the yield stress,[45] and its value closely matches that of the critical stress associated with viscosity bifurcation.[35] In addition, some thixotropic systems exhibit delayed yielding, wherein viscosity initially grows toward divergence but then drops to a constant value describing flow as the material yields. Such behavior has been attributed to the alteration of the relaxation time spectrum by the deformation field.[37, 40] These aspects introduce an additional layer of complexity in defining yield stress, its time dependence, and especially what kind of deformation material undergoes for values of stresses lower than the yield stress. For non-thixotropic yield stress materials, where behavior shows Bingham-type response, some materials may undergo localized plastic deformation before complete structural breakdown rather than transitioning from a solid-like state to fully viscous flow.

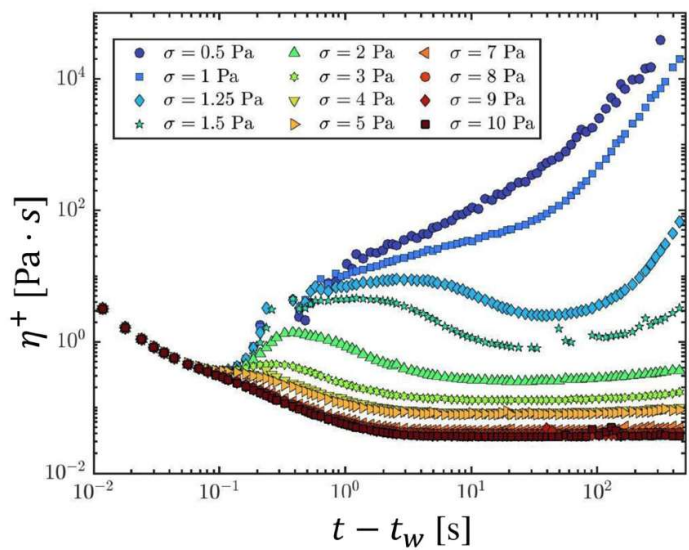

**Figure 2:** Demonstration of a viscosity bifurcation in bentonite dispersion at a specific aging time when subjected to various constant stresses. The plot shows the evolution of transient viscosity ($\eta^+$) as a function of creep time. For stresses below a critical value (between 1.5 and 2 Pa, but it depends on aging time), viscosity continues to grow with diminishing shear rate, eventually arresting the flow. In contrast, application of stresses above the critical value induces a steady-state flow with constant viscosity. Reproduced from ref 43. Available under a CC-BY 3.0 license. Copyright The Royal Society of Chemistry Rathinaraj *et al.*[43]

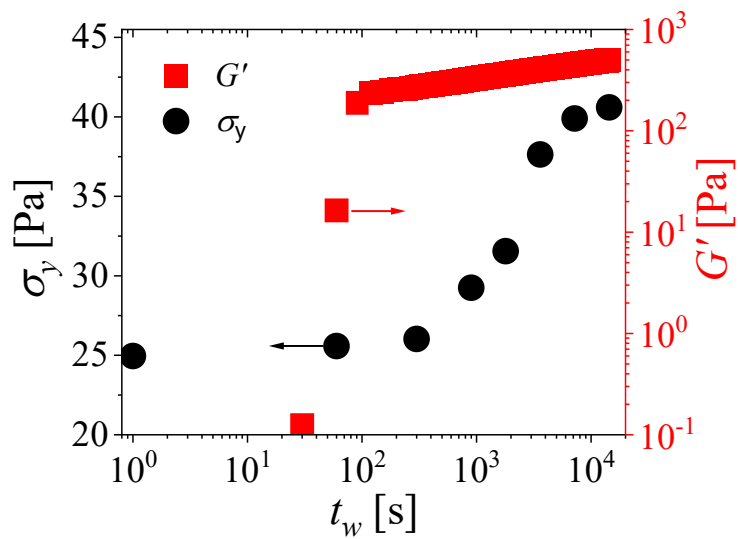

**Figure 3:** Time (age) dependent evolution of yield stress and storage modulus is shown for an aqueous dispersion of clay. Since cessation of shear melting, yield stress remains

constant for a particular duration but subsequently shows a power law increase with aging time. Data is from Bhattacharyya *et al.*[35]

Yield stress is further divided into two main categories: Static and Dynamic Yield Stress. Static yield stress refers to the minimum stress required to induce flow in a material, which is under quiescent conditions. Static yield stress is often influenced by the time elapsed since cessation of shear melting or the time over which it has remained undisturbed. Dynamic yield stress, on the other hand, corresponds to the minimum stress required to maintain flow once it has started. This distinction becomes particularly relevant in thixotropic materials, where physical aging or structural build-up during rest causes an increase in static yield stress as shown in Fig 3. For thixotropic materials the static yield stress usually forms an upper bound for the dynamic yield stress. Rheological protocols such as stress ramp-up versus ramp-down,[48-49] creep tests after different rest periods,[37] or shear start-up[28] are commonly used to highlight this disparity. Failure to recognize this distinction may lead to inconsistencies in yield stress determination and misinterpretation of flow behavior. Notably, in some thixotropic systems that display an increase in modulus as shown in Figs. 1 and 3 and hyperaging dynamics, wherein relaxation time evolves stronger than linearly with respect to time,[50] the flow curves can become non-monotonic, especially at low shear rates, where steady state stress decreases with an increase in shear rate.[11, 35] Such inverse dependence between shear stress and shear rate makes the system unstable, rendering uniform flow impossible. Consequently, the flow field shows steady-state shear banding, wherein flowing and non-flowing bands coexist under the same applied stress.[51-55] Such behavior is usually reported for hyperaging materials that also show dynamic yield stress. The presence of shear bands challenges conventional interpretations of yield stress as it suggests a more complex underlying rheophysical phenomenon.

Accurately measuring the yield stress in soft materials also presents several experimental challenges. The inherently transient nature of structure evolution means that yield stress values depend on the experimental protocol used for measurement. Since structure evolves dynamically, measured yield stress values often depend on the timescale of the experiment. Consequently, performing a steady shear experiment over a limited time usually gives a false impression of finite viscosity, thereby showing an erroneous Newtonian plateau in the limit of small shear rates.[56-57] This issue is particularly critical in weak gels and emulsions, where yield stress values may be close to the sensitivity limit of the rheometer. The precise value depends on the experimental protocol used, even in materials with well-defined yield stress, such as highly filled polymer melts and concentrated colloidal suspensions in glassy state. A major source of ambiguity also arises because, many times, yield stress is not directly measured but instead extrapolated from experimental data using empirical models. Different extrapolation techniques lead to different values. This complicates comparisons across studies. In addition, in practical rheometry, obtaining reliable yield stress values below 0.01 Pa is difficult, despite some reported values being lower. The primary physical limitation to extremely low yield stresses is Brownian motion, which disrupts weak colloidal structures, preventing them from maintaining rigidity over long timescales.[7, 23]

Bingham and Herschel-Bulkley models well capture the non-thixotropic yield stress behavior.[24, 28] However, for thixotropic materials that show yield stress, advanced rheological models have been developed that incorporate combinations of elastic and viscous elements, often complemented by a structural parameter, which is a conceptual measure of microstructure that evolves dynamically with stress and time.[17, 58] These parameters are governed by kinetic equations describing structural breakdown (rejuvenation) and build-up (aging), enabling the models to capture time-dependent yield stress, hysteresis, and viscosity bifurcation.[10, 12, 35, 48] Importantly, many such models, often termed thixo-visco-plastic or thixo-elasto-visco-plastic,[29, 58-62] treat the yield point as a fluidity transition, describing divergence of viscosity, rather than a sharp elastic-plastic boundary. Interestingly, some of these models do include a conceptual slider element that incorporates the yielding term as per Bingham's original framework. However, the actual flow is in those models is still governed by a balance between structural evolution and how imposed deformation affects the same. These models are sometimes termed as "plastic." However, these are fundamentally built to describe flow governed by evolving internal structure, rather than plastic deformation in the classical solid mechanics sense. As discussed in subsequent sections, making a distinction between such "visco-plastic" flow and true plastic deformation is critical for developing unified rheological frameworks, which bridge the soft matter and solid mechanics paradigms.

This discussion suggests that the yield stress fluids cannot be reduced to a single threshold stress value. Instead, yielding encompasses a broad spectrum of behaviours, including elastic, viscous, and plastic responses. Notably, thixotropic materials introduce further complexity through time-dependent yield stress, viscosity bifurcation, and hysteresis, posing challenges to the original definition of yield stress proposed by Bingham. Particularly, what kind of deformation a material undergoes before the so-called yield stress threshold is a key question that needs further investigation and deliberation.

### 3. Plasticity

Historically, in rheology as well as in the solid mechanics literature, the term plasticity has carried two distinct meanings. In the former, following Bingham,[1-2] plasticity refers to the flow of yield-stress fluids beyond the yield point, suggesting an unbounded, irreversible continuous increase in strain under application of constant stress. On the other hand, in solid mechanics, it represents irreversible constant strain after the yield stress threshold is exceeded.[63] Accordingly, in rheology, plasticity implies rate-dependent (continuous) and homogeneous flow. On the other hand, in solid mechanics, it indicates finite, rate-independent, constant irreversible strain that may not be homogeneous. Assessment of this discrepancy is the subject of the subsequent discussion, wherein we emphasize how the rheological usage diverges from the classical solid mechanics perspective.

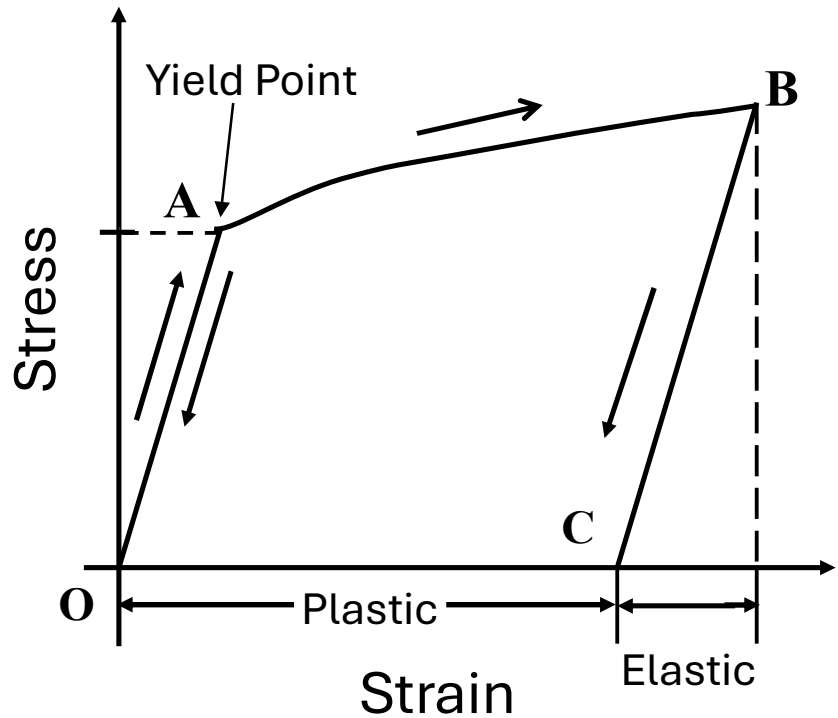

**Figure 4:** Stretching of metals and plastics associated with necking.

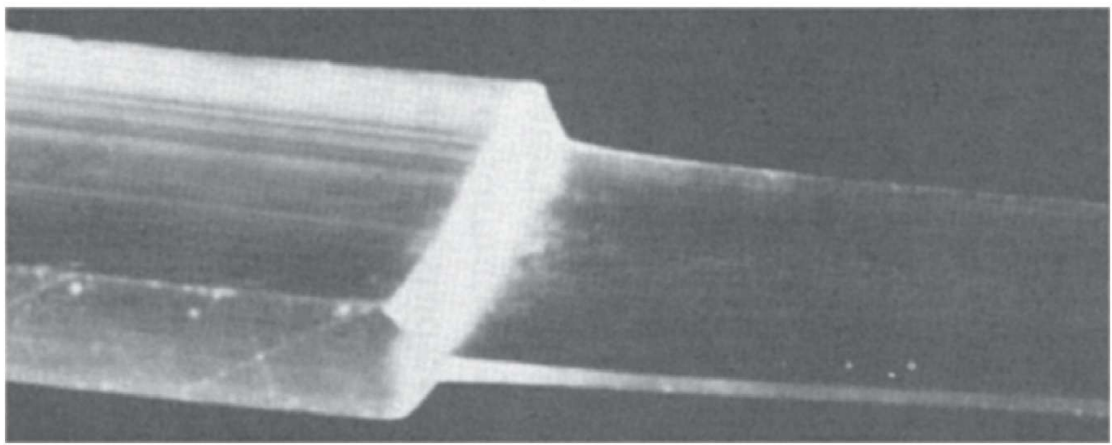

**Figure 5:** Necking during extension of polyethylene bar. The figure is reproduced with permission from Ward and Sweeney.[64] Copyright 2012 John Wiley & Sons, Ltd.

In solid mechanics, on the contrary, plasticity is understood differently. It refers to irreversible deformation in elastic solids that occurs after surpassing a certain stress threshold ("yield point"), typically at large strains.[63] A classic example of this phenomenon is the stretching of "mild" metals such as low-carbon steel or copper, illustrated in Fig. 4. On the stress-strain curve, when strain induced in the material exceeds the strain associated with the yield point A, irreversible strain begins to accumulate. If the stress on the material is unloaded at point B, the corresponding path follows line BC, wherein 0C represents the irreversible plastic strain. Significantly, this plastic deformation is not associated with flow. A similar stress–strain curve is observed during the tensile deformation of solid polymers. As with metals, this behavior of plastics is often accompanied by necking, as shown in Fig. 5. In semicrystalline polymers such as polyethylene, this necking corresponds to cold drawing, i.e., localized plastic deformation with neck propagation. The deformations associated with neck formation are irreversible and are often referred to as plasticity, particularly in engineering literature.[64-65] The plasticity of solids enables key industrial processes such as forging, rolling, and stamping of metal components. However, plastic deformation can also be undesirable, such as in the deformation of automobile body parts during collisions. Throughout, we use the analogy to metals only as a conceptual bridge to emphasize rate-independent, finite plastic strain; the microscopic carriers of plasticity in semicrystalline polymers (chain/lamella processes) are not mechanistically equivalent to dislocation-mediated crystal plasticity in metals.

The onset of plastic flow in isotropic materials must remain unaffected under coordinate transformations. The yield point, therefore, must depend only on the deviatoric Cauchy stress tensor ($\boldsymbol{\sigma}$) expressed through its invariants ($II$) and ($III$):[63]

$$F[f_1(II), f_2(III)] = 0, \qquad (1)$$

where $II = tr(\boldsymbol{\sigma}^2)$ and $III = tr(\boldsymbol{\sigma}^3)$. Among various possible forms, the von Mises yield criterion is widely used for both crystalline, amorphous as well as soft materials. In terms of components of a stress tensor, it reads:[63]

$$\sigma_y^{*2} = \frac{1}{2}[(\sigma_{11} - \sigma_{22})^2 + (\sigma_{22} - \sigma_{33})^2 + (\sigma_{33} - \sigma_{11})^2 + 6(\sigma_{23}^2 + \sigma_{31}^2 + \sigma_{12}^2)], \qquad (2)$$

where $\sigma_y^*$ represents the von Mises condition of yielding and $\sigma_{ij}$ correspond to various components of the stress tensor. This condition corresponds to the energy required for isochoric (volume-preserving) shape change and has found success not only in modeling metal plasticity but also in describing yielding in yield-stress fluids and soft glassy materials.[8-9]

Before the yield point, a material undergoes elastic deformation given by:

$$\sigma_{12} = G\gamma_{el}, \qquad (3)$$

where $G$ is modulus and $\gamma_{el}$ is elastic strain. Beyond the yield point, an increase in applied stress induces plastic deformation in soft materials, which leads to an important parameter called modulus of plasticity ($P$) given by:[63]

$$\sigma_{12} = P(\gamma_{pl})\gamma_{pl}, \qquad (4)$$

where $\gamma_{pl}$ is the accumulated plastic shear strain while $P$ represents the resistance offered by the material after yielding to irreversible deformation. It is important to note that this relationship, like Hooke's law, does not include the time factor. Very interestingly, based on experimental observations, Kachanov[63] reported that: $0 < P(\gamma_{pl}) < G$. The plastic modulus provides a physically realizable indicator of the material's stiffness in the plastic regime. A higher plastic modulus suggests greater hardening of the material, and hence it deforms less easily after yielding.[32] It should be noted that there is a direct analogy between the plastic modulus and the elastic modulus. Both can be regarded as material constants in the case of linear behavior of the medium. More generally, however, both depend on the deformation history and the material's structure, and are therefore better viewed as material functions rather than true constants.[32]

## 4. Plastic Deformation in Soft Materials

We now discuss some important cases and experimental observations associated with plastic deformation arising from different microscopic processes. Metals are mentioned only as a backdrop for comparison. As emphasized before, the fundamental difference between flow and plasticity gets exemplified through a simple shear experiment in which a constant stress is applied (see Fig. 6). It should be noted that both viscous flow and plastic deformation are irreversible, therefore, once the stress is removed, the material does not revert back to its original state although partial elastic recovery may occur upon unloading. However, the key difference lies in the nature of deformation under constant stress. In liquids, the application of constant stress causes viscous strain to increase continuously with time. In contrast, when plastic solids are subjected to the same constant stress, the irreversible strain induced in the same is constant and does not evolve with time. This difference, illustrated in Fig. 6, forms the basis for a general approach to defining plasticity in rheology, particularly for isotropic media. If we consider the process of irreversible deformation to be quasi-equilibrium in nature, then the relationship between stress and strain should not involve time explicitly, that is, it should be independent of the deformation rate. Following this logic, the formal description of plastic deformation should be structurally similar to that of elastic deformation, i.e., represented as a direct relationship between stress and strain, particularly for changes in shape.

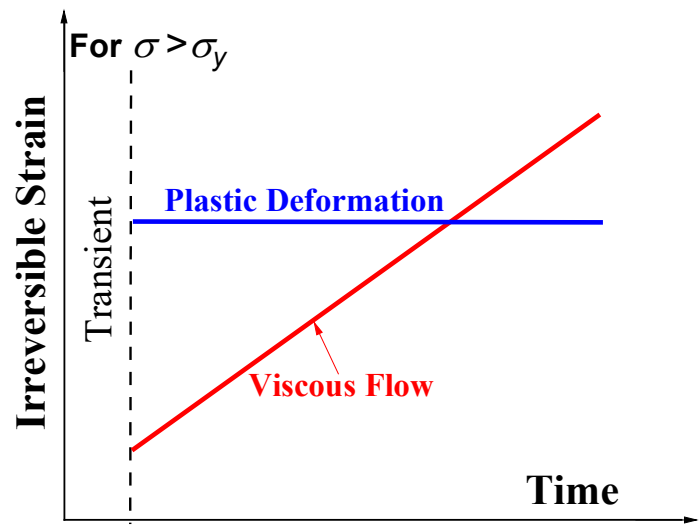

**Figure 6:** Development of irreversible deformations in viscous flow or plastic deformation under application of constant stress beyond a yield point. The area to the left of the vertical line corresponds to the transient mode of deformations (flow, elastic, or viscoelastic).

Modelling plastic behavior in soft materials is naturally based on principles traditionally applied to metals and crystalline solids. However, such efforts require adaptation to account for the structurally arrested, often amorphous microstructures encountered in soft materials. In classical plasticity, constitutive laws are typically formulated using stress–strain relationships that are rate-independent and history-sensitive, such as Prandtl–Reuss model.[66] However, in soft materials like dense suspensions, colloidal gels, pasty materials, or polymers, plastic deformation may be governed by localized particle rearrangements, network rupture, or entanglement slippage. As such, phenomenological models belonging to a family of structural kinetic models have emerged that incorporate internal state variables, such as accumulated plastic strain or evolving structural rigidity, into the constitutive framework.[17, 58-59] However, unlike flow models where time or strain rate directly governs stress, plasticity models in soft systems emphasize the path-dependent evolution of stress-free configurations, allowing them to distinguish between transient flow and genuine plastic deformation. Nonetheless, in many soft systems, both rate-dependent and path-dependent behaviors could be present, with one followed by the other, and hybrid models may be required to fully capture their mechanical response.[59, 61, 67]

Interestingly some soft materials may exhibit strain hardening or structural evolution, wherein microstructural rearrangements upon deformation may cause P to increase. This conceptualization of $P$ allows for treating plasticity not merely as a post-yield phenomenon but as a distinct constitutive behavior, characterized by an effective modulus in the same way as elastic modulus and viscosity are, thereby making it amenable to continuum modelling and practical material design.[63, 66] It should also be noted that many real materials show some rate dependence, as the slip events could be weakly rate-dependent, in certain materials. This means the rate at which the material is deformed may affect how it flows plastically. In rheological terms, although such rate dependence appears when analyzed at the bulk level and may appear similar to what is predicted by the Herschel-Bulkley equation, in which stress is expressed as a rate-dependent parameter. However, as mentioned, plasticity describes inhomogeneous, localized deformation, which is governed by accumulated strain. On the other hand, models like Herschel–Bulkley describe a homogenized, continuum flow regime, where rate governs stress response. These frameworks reflect two ends of a spectrum and, therefore, are not directly interchangeable.

It should be noted that the elastic-to-plastic transition is a nonlinear phenomenon and is typically accompanied by structural transformations. In metals, plasticity is primarily associated with alteration in the crystalline structure. Two principal mechanisms are usually considered to govern the irreversible structural transformation: slip and twinning. During the slip, the crystal blocks slide over one another along specific crystallographic planes. Twinning, on the other hand, involves a reorientation of a portion of the crystal lattice so that it becomes symmetrically aligned with the untwined portion of the lattice.[68] Similarly, the plasticity observed in solid polymers is linked to transformations in their super-crystalline structures, as well as to the orientation and alignment of polymer chains under deformation. However, it should be emphasized that the nature of plastic deformation in metals and polymers is fundamentally different. In the former, plasticity arises primarily from crystal deformation mechanisms, which include dislocation motion as well as phase transformations. On the other hand, the plastic response in polymeric systems may involve phenomena like cold drawing, incorporating far more complex processes, such as chain stretching, lamellar slip, crystalline reorientation, and structural anisotropy, as discussed below.

Before analysing the real experimental data on plastic behavior, let us first differentiate between three types of responses: viscoelastic materials, visco-elasto-plastic materials, and elasto-plastic materials. The creep-recovery responses (i.e., a constant stress is applied for a certain duration followed by a recovery period under zero stress) of these three types of materials are illustrated in Fig. 7. Fig. 7(a) shows the response of a standard viscoelastic material, wherein after an initial transient, the material flows at a constant shear rate. Upon removal of stress, the material undergoes partial recovery, which is characteristic of viscoelastic behavior. In Figs. 7(b) and 7(c) we schematically represent the responses of visco-elasto-plastic and elasto-plastic materials, respectively. The primary difference between these two responses lies in their behavior prior to yielding. In the former case (visco-elasto-plastic), the material behaves as a viscoelastic solid, typically represented by a model such as Kelvin-Voigt. On the other hand, the latter response (elasto-plastic) indicates a purely elastic response before yielding.

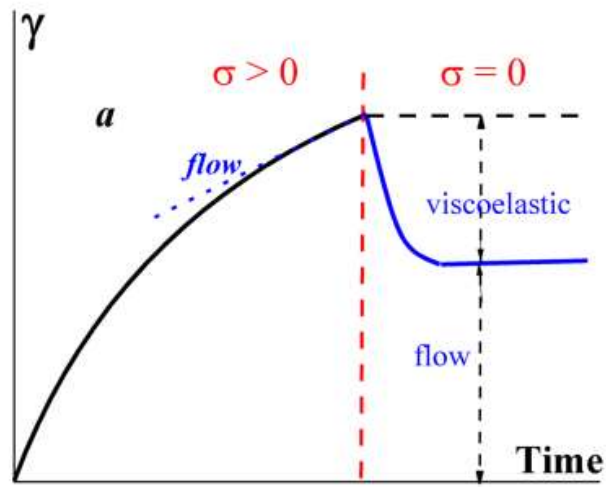

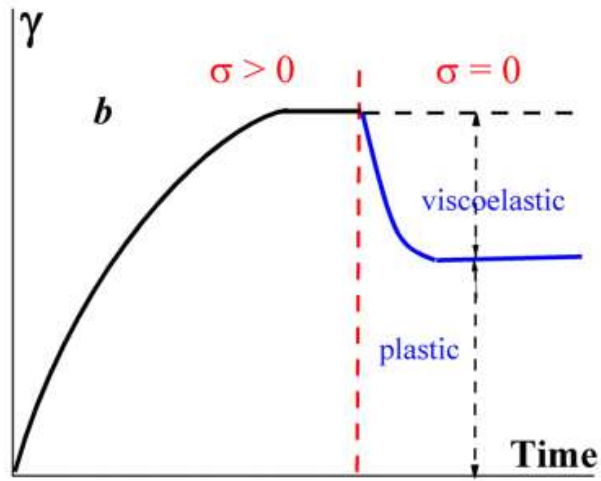

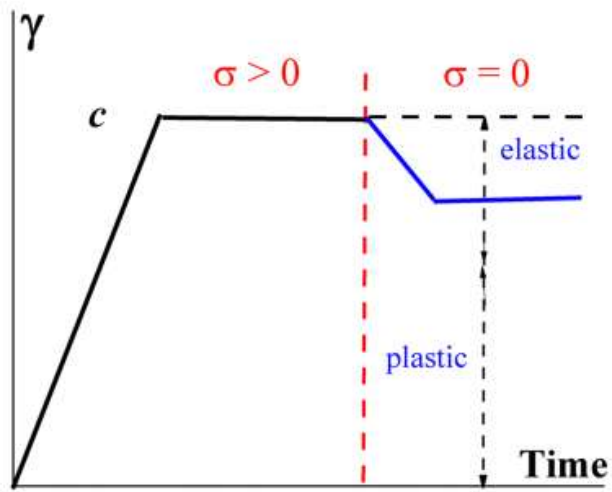

**Figure 7:** Three types of rheological behavior in creep: viscoelastic fluid (a), visco-elasto-plastic solid (b), and elasto-plastic medium (c)

The experimental observations on 60 volume % suspension of Al powder in polyethylene glycol by Malkin and coworkers[69-70] revealed a clear manifestation of true plastic behavior, as shown in Fig. 8. It can be seen that in the creep experiment the strain induced by any applied stress remains constant over time, and this strain increases progressively with increasing stress. The elastic strain associated with this total strain has been reported to be 40 %, and this fraction is nearly independent of stress. This suggests 60 % of the induced strain is purely plastic in nature.[69-70] The authors report similar behavior for concentrations beyond the jamming transition and also for a bi-disperse particulate system.[69-70] The studied dense suspensions, as per rheological classifications, behave as solid-like material. While the dispersed solid particles are rigid and resist deformation elastically, the jammed system can also undergo finite, irreversible plastic strain under higher loads.

In densely packed particulate suspensions, the yield stress arises due to jamming of particulate matter, whose mechanically self-supporting configuration resists deformation. Material yields when the applied stress overcomes the resistance offered by the jammed constituents, unlocking the system from its arrested configuration. However, in the systems studied, when stress exceeds the yield stress, it does not result in a freely flowing continuum. Instead, the particulate configuration undergoes localized rearrangements, seeking new metastable conformations that can support the higher imposed stress. This process is inherently intermittent and collective: with a gradual increase in applied stress, particles unjam only to reconfigure into new jammed states capable of resisting the next increment of stress. With an increase in the applied stress, the jammed particulate network accommodates it through further plastic deformation through additional irreversible particle rearrangements. As stress builds up, it induces further structural changes in the material, which causes plastic strain to build up over time.  As a result, the microstructure evolves with every increment in stress, with higher magnitudes causing more extensive rearrangements and, consequently, greater plastic strains. This behavior distinguishes such suspensions from simple Bingham-type models: the flow is not continuous but mediated by a cascade of structurally adaptive jamming–unjamming transitions. In this regime, the increase in plastic strain with stress reflects not a steady flow, but a sequence of particle-scale reorganizations driven by the system's attempt to maintain mechanical stability under growing load.

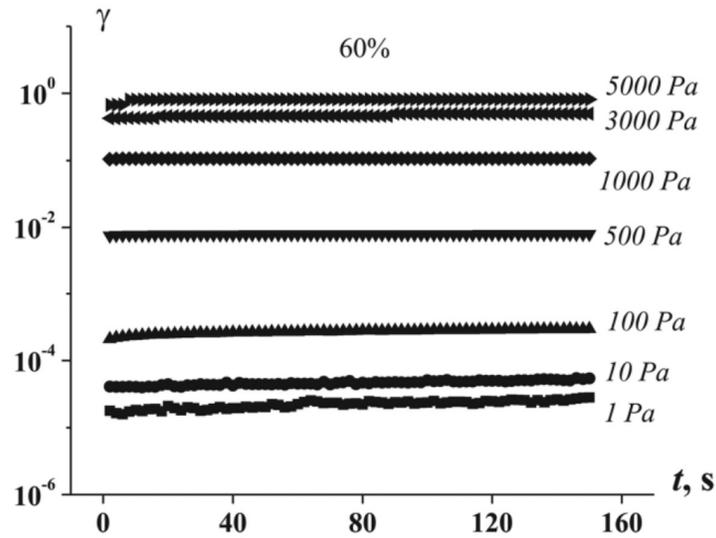

**Figure 8:** Creep behavior of suspension of Al powder in polyethylene-glycol (concentration 60 vol. %) under different shear stresses. Reproduced with permission from Malkin et al.[69] Copyright 2020 the Society of Rheology.

Plastic deformation is also characteristic of ultra-high molecular weight polyethylene (UHMWPE). Under tensile loading at ambient temperatures, UHMWPE primarily undergoes cold drawing, which is a form of plastic deformation. During cold drawing strain localizes and then propagates via necking, consistent with the plateau strains seen in Fig. 9.[71] In polymers such as UHMWPE, a viscoelastic transition region typically precedes the onset of ultimate plastic deformation. As shown in Fig. 10, the dependence of the plastic modulus on stress for UHMWPE differs notably from one sample to another. The qualitative difference in the stress-dependence of $P$, including its magnitude, in three systems is attributed to differences in molecular architecture, particularly the presence of lower molecular weight chains that significantly influence plastic behavior.

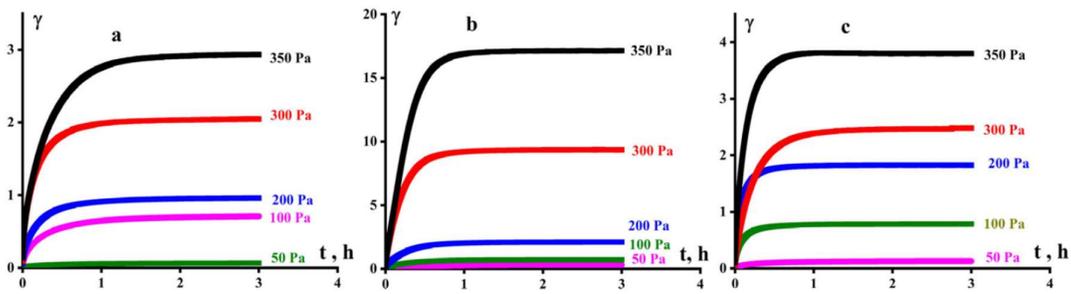

**Figure 9:** Creep behavior for three samples of UHMWPE from different sources (a) GUR 4150 (Mw=$9 \times 10^6$ Da), (b) GUR 4120 (Mw=$4 \times 10^6$ Da), and (c) UTEC 6540 (Mw=$8 \times 10^6$ Da). Reproduced from ref 71. Available under a CC-BY 4.0 license. Copyright MDPI Malkin *et al.*[71]

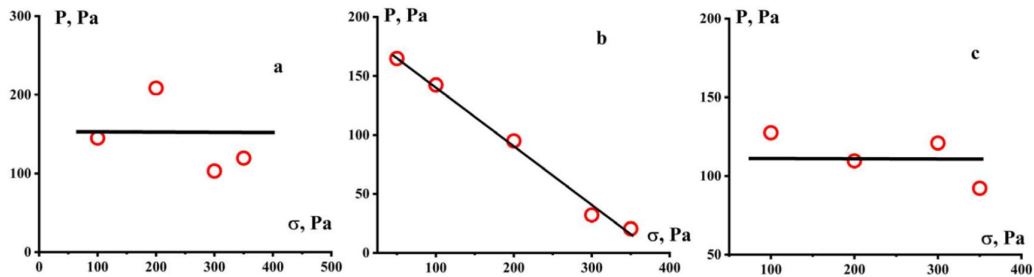

**Figure 10:** Modulus of plasticity for three samples of UHMWPE (a) GUR 4150, (b) GUR 4120 and (c) UTEC 6540. Reproduced from ref 71. Available under a CC-BY 4.0 license. Copyright MDPI Malkin *et al.*[71]

In semicrystalline polymers in the leathery/rubbery-amorphous regime, plastic response arises from a coupled set of processes, amorphous-phase chain stretching, lamellar slip and rotation, crystalline reorientation, and tie-chain stretching, which produce a strong anisotropy along the draw direction. It is known that UHMWPE constitutes two phase semicrystalline structure consisting of stacks of orthorhombic lamellae and amorphous layers that are densely entangled and threaded by tie chains. Below the melting temperature, both crystalline slip as well as entanglement disengagement dynamics are not expected on experimental time–scales.[64, 72] Under the application of constant stress, the plastic deformation (cold drawing) includes interlamellar shear (chain slip) along crystallographic planes,[73] which is followed by Lamellar rotation and fragmentation. Eventually, the tie chains undergo stretching wherein the amorphous strands linking neighboring lamellae extend affinely with macroscopic strain. Their deformation of tie chais is entropy dominated, but at temperatures significantly smaller than the melting point, the entropic force is effectively "frozen in" as schematically shown in Fig. 11. The tie chains cannot relax on practical timescales as reptation times for UHMWPE chains are orders of magnitude longer than the experimental timescale. The microstructure is therefore locked into the deformed state, giving genuine plastic (irrecoverable) strain even though the underlying mechanism is chain straightening. Since the tie chain tension rises nearly linearly with chain stretch, the net incremental stress needed to create an incremental plastic strain is constant, leading to a constant plasticity modulus.

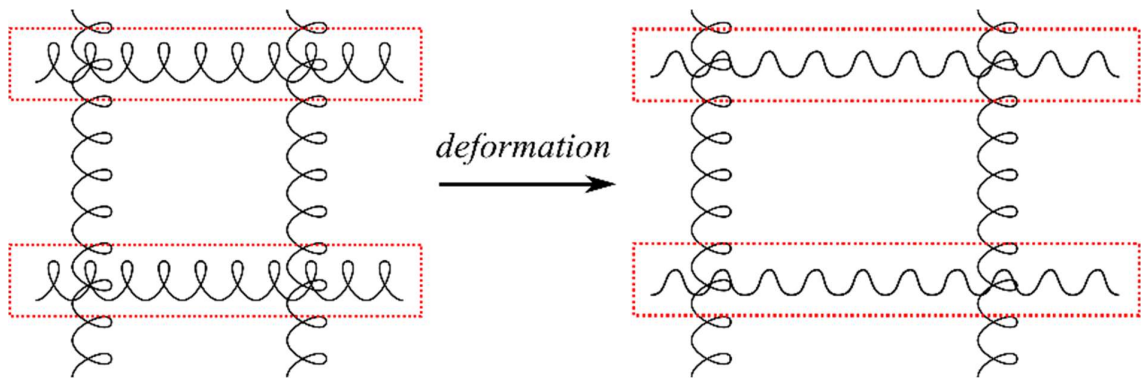

**Figure 11:** Diagram illustrating the cold-drawing plasticity mechanism of UHMWPE

In addition to the above two examples of dense particulate dispersion and semicrystalline polymer, several soft consumer materials may also display true plastic behavior. These materials have soft solid like consistency and may exhibit rate-independent, irreversible deformation under applied stress, consistent with the definition of plasticity used in solid mechanics. Possible examples may include bar soaps, [74] chocolates bars,[75] processed cheeses,[76] wax-based formulations such as lip balm/stick,[77] deodorant sticks, etc. Wax based formulations, chocolate bars, and bathing soaps may undergo localized yielding under applied stress, often leading to permanent shape change, which is an important process associated with stamping. Similarly, processed cheese and butter deform plastically when spread, with irreversible shear strain that does not fully relax once stress is removed. These examples underline the importance of distinguishing plasticity from homogeneous viscous flow in soft materials. Their mechanical response may not be fully represented by conventional rheological models such as Bingham or Herschel-Bulkley and may require frameworks that incorporate local plastic deformation, and strain-hardening, similar to those used in metals or polymers. More experimental and theoretical work is needed to understand plastic deformation in such high-volume soft consumer products. To consolidate these distinctions, Table 1 gives a summary of different definitions and physical origins of plasticity and yielding in Solid Mechanics and Rheology.

**Table 1.** Traditional use of the term plasticity in mechanics and rheology

| Aspect | Metals / Solid Mechanics (Hard Matter, Polymers) | Rheology / Soft Matter (Yield Stress Fluids, Gels, etc.) | Remarks / Clarifications |
|---|---|---|---|
| **Definition** | Irreversible, finite, *rate-independent* deformation after the yield point; involves a structural transition within the solid. | Often used synonymously with *flow* beyond a yield stress, it implies continuous, irreversible deformation that is typically *rate-dependent*. | The term "plastic flow" in rheology diverges significantly from classical plasticity in solids. |
| **Physical carriers of plasticity** | Dislocation motion, slip, twinning, and crystal reorientation; the crystal lattice is deformed. In polymeric systems, lamellar slip, chain orientation, and tie-chain stretching | Localized particle rearrangements, shear transformation zones (STZs), and network rupture. | Metals: crystalline defects; Soft matter: microstructural rearrangements (amorphous, semicrystalline, colloidal). |
| **Yield criterion** | Based on invariants of the deviatoric stress tensor (e.g., von Mises, Tresca), it defines a stress threshold beyond which plastic deformation begins. | "Yield stress" is often defined as the stress above which flow occurs (Bingham, Herschel–Bulkley models). For thixotropic materials, static vs. dynamic yield stress differs. | Rheological "yield stress" differs from the solid mechanics yield stress; |

23 | P a g e

| | | | |
|---|---|---|---|
| **Stress–strain response** | Stress–strain curve shows elastic region up to yield point; plastic region exhibits hardening, softening, or necking (rate-independent). | Creep/recovery: viscoelastic or visco-elasto-plastic materials show rate-dependent strain accumulation; thixotropy can cause viscosity bifurcation. | Metals: plateau/strain hardening; Soft matter: time-dependent recovery/aging complicates clear demarcation. |
| **Role of time** | Plastic strain is permanent and does not evolve with time under constant stress. Plastic-like plateaus may occur in jammed systems (e.g., UHMWPE, concentrated suspensions) | Viscous strain increases continuously with time under constant stress. | Fundamental distinction: solids → time-independent; soft matter → often time-dependent. |
| **Examples** | Mild steels, copper, aluminium (crystal plasticity). Necking if stretching solid plastics | Colloidal gels, concentrated emulsions, dense particulate suspensions | Illustrates the diversity of plastic carriers in soft matter. |
| **Terminology note** | "Plastic deformation" is solid-to-solid transition | "Plasticity is solid-to liquid transition | This terminological overlap is a major source of confusion |
| **Conclusion** | Plasticity of solids and yielding of liquids are two different phenomena. They can exist independently (both or just one of them) in the same soft matter | | |

## 5. Microscopic Origins of Plasticity in Soft Matter

Figs. 8 and 9 represent the macroscopic stress–strain behavior during plastic deformation. This behavior is rooted in microscopic, localized plastic events. Such events have also been identified in soft materials, although their clear manifestation at the bulk level may not always be captured as a response, as presented in Fig. 8 or 9. Notably, recent advances in experimental techniques such as particle-scale imaging and scattering, as well as computer simulations, have revealed that soft jammed materials exhibit localized plastic events at the microscopic scale. These events bear a striking resemblance to the plastic deformation mechanisms observed in crystalline metals. Such observations challenge the traditional view that flow in soft materials is always homogeneous and well-described by Bingham or Herschel-Bulkley models. These findings also provide a microscopic link between the continuum rheology and the solid-mechanics view of local plastic deformation prescribed by classical metallurgy. Stress induces irreversible, localized particle rearrangements in soft jammed materials, where the plasticity originates in these systems, which are analogous to dislocation motion in metals.[78-80] In soft jammed materials, at a local yielding event, particles' (or jammed elements) displacement deviates from the expected affine deformation, leading to localized plastic zones. these regions resemble the correlated plastic rearrangements observed in theories of shear transformation zones (STZ) in amorphous solids.[81-82] Ridout and coworkers[80] simulated athermal quasistatic shear of jammed packings and identified structural features, such as softness and low coordination number, that strongly correlated with sites of future plastic rearrangements, in a manner reminiscent of defects governing yielding in crystalline solids. Falk and Langer[83] showed that localized plastic rearrangement of particles produced a fixed irreversible strain increment, which was independent of strain rate, confirming its rate-independent, plastic character. Patinet et al.[84] reported a strong correlation between local yield stress and plastic rearrangements in amorphous solids. They concluded that the local yield stress is a reliable predictor of deformation sites, even after multiple plastic rearrangements. Recently, Gury et al.[85] showed that local yielding events are strongly influenced by the microstructure of a soft material, which results in distinct, microstructure-dependent macroscopic yielding and post-yielding rheological behavior.

Interestingly, these plastic rearrangements in soft materials are not isolated; but often occur in intermittent bursts or avalanches, reflecting collective, correlated motion of particles, which are reminiscent of slip events and shear banding in metallic glasses.[86-87] Interestingly, Maloney and Lemaître[87] reported similar results for athermal systems, suggesting that thermally activated motion is not a prerequisite for STZ formation in soft matter, especially for systems like dense emulsions or compressed granular materials. Localized shear transformations redistribute stress elastically over long ranges and hence may trigger cascades of subsequent rearrangements. Simulations of elastoplastic models reveal that the slip avalanches follow power-law size distributions with cutoffs that diverge as the system approaches the macroscopic yielding point.[88] 3D particle dynamics simulations to analyze yielding behavior in soft particle revealed that the nonaffine dynamics at the particle scale, including transient caging and localized rearrangements, are crucial for the onset of flow.[89] This showed that localized plastic events at the particle level are central to the yielding of soft microgel glasses. Masschaele et al.[90] employed video microscopy and directly observed localized bond breakages in a 2D colloidal gel network subjected to shear. They reported that these break-up events are shown to initiate a cascade of further events, ultimately leading to macroscopic yielding. Confocal microscopy of hard-sphere colloidal glasses under shear performed by Schall et al.[91] revealed striking similarities between localized particle rearrangements in colloidal glasses and shear bands in ductile metals. These micro-shear bands form connected networks along planes of maximum resolved shear, serving as precursors to macroscopic yielding. Thus, the mechanics of plastic flow in soft jammed systems bear deep parallels with solid mechanics.

Building on these microscopic insights, elastoplastic lattice models facilitate coarse-graining of the material into elements that accumulate elastic strain until an STZ-like yield criterion. This triggers localized yielding events leading to the redistribution of stress. Such models connect the non-local rheology of dense suspensions and the size-dependent strength of amorphous solids. Interestingly, their governing equations do not contain viscosity explicitly, and the macroscopic flow emerges from the spatiotemporal statistics of discrete plastic rearrangements. This provides a natural bridge between continuum visco-plastic flow laws and classical (rate-independent) plasticity.[88] This discussion also suggests that many "yield-stress fluids" may undergo locally rate-independent plastic deformation events, even though the ensemble-averaged response of a cascade of such events is well described by Bingham or Herschel–Bulkley-type laws. The apparent viscosity in such materials could, therefore, originate from an averaged quantity reflecting the density and interaction of STZs and avalanches. Incorporating this physics into continuum rheological models is a challenging task. Nonetheless, efforts have been made to propose kinetic theories for soft glassy materials that incorporate the rate and spatial distribution of plastic events, that link microscopic plastic rearrangements and macroscopic flow via a coarse-grained constitutive law.[78] A recent perspective by Divoux et al.[92] presents a more comprehensive discussion on the correlation between local pastic events and macroscopic yielding.

Finally, in this discussion on plasticity, it is important to emphasize that this phenomenon, long recognized and widely exploited in the context of metals and ceramics, may also extend to a variety of soft materials. The localized plastic events in many materials may lead to eventual homogeneous flow beyond mesoscopic length scales, showing validation of Bingham-type constitutive response. However, over the bulk scale, such flow beyond yielding still needs to be characterised as a viscous response, although its initiation might have taken place through localised plastic events. A distinction should be made when applied stress leads to plastic events that lock the microstructure, thereby rendering the system rate-independent permanent deformation, leading to a plasticity modulus reminiscent of that observed in metals. The motivation of this article is essentially to clarify that at the macroscopic level, these two behaviors are different, and one should keep this aspect in mind while studying this very intriguing class of materials.

## 6. Conclusions

A broad class of structured materials, from soft solids such as colloidal and polymeric systems to metals, undergo elastic deformation below a threshold stress, often termed yield stress. How material deforms beyond the yield point is a matter of intense investigation over the past several decades. Various disciplines term the deformation beyond the yield point to be plastic, but there happens to be a significant difference among various classes of materials. The question is whether such deformation should be termed as viscous or plastic. Interestingly, the term "plasticity" has evolved independently within two disciplines, rheology and solid mechanics, leading to a divergence in interpretation. In rheology, the term plastic flow is used for continuous, irreversible deformation (i.e., viscous flow) beyond the yield point as modelled in Bingham or Herschel-Bulkley constitutive equations. Importantly, in rheology, the plastic flow is often rate-dependent and is governed by viscosity. Classical plasticity, on the other hand, refers to finite, rate-independent, irreversible deformation that occurs in solids beyond the yield stress threshold. The fundamental distinction lies in the nature of the deformation: rheological plasticity describes flow, whereas classical plasticity describes a structural transition to a permanently deformed solid. Accordingly, references to metal plasticity here serve as conceptual scaffolding for distinguishing rate-independent plasticity from viscous flow; they are not intended to imply mechanistic identity in soft, semicrystalline, or jammed systems.

In soft materials, especially structurally arrested or jammed systems, both types of behavior may coexist; therefore, it is essential to have clarity. For instance, densely packed suspensions or ultra-high molecular weight polymers exhibit true plastic deformation with rate-independent strain plateaus under constant stress, akin to metals and crystalline solids. On the other hand, these materials are often modeled using visco-plastic frameworks that do not distinguish between flow and true plastic deformation. This conflation becomes especially problematic in materials exhibiting thixotropy, where the apparent yield stress is a dynamic property determined by the kinetics of inherent physical aging and deformation field-induced rejuvenation behaviors, rather than a sharp, intrinsic material threshold.

In thixotropic materials, the stress threshold, below which viscosity diverges, is associated with viscosity bifurcation and is often labelled yield stress without the deformation necessarily being elastic below the threshold. Additionally, in some thixotropic materials, the stress at which flow initiates (static yield stress) may differ substantially from the stress required to sustain flow (dynamic yield stress), especially when structural recovery processes are at play. The yield stress may also show time dependence, and this overall phenomenon introduces ambiguity in both defining and measuring yield stress. This work emphasizes the need for conceptual clarity in defining yield stress and plasticity in soft materials. Not all irreversible deformation following a yield point in soft jammed materials constitutes either homogeneous rate-dependent viscous flow or rate-independent classical plastic deformation. Recognizing this distinction is important for developing accurate constitutive models and interpreting experimental data.

Ultimately, this article tries to distinguish and clarify the use of different terminologies and frameworks that respect the physical origins of the mentioned deformation processes. By elaborating on differences between classical plasticity concepts and those used in conventional Bingham-type approaches in rheology, we aim to bridge the gap between soft matter physics and solid mechanics. Accordingly, the discussion of metals should be viewed as a conceptual scaffold, and not as a co-focus. The central purpose of this article is to reassess plasticity and yielding in soft materials. We believe this clarification will enhance theoretical understanding and inform material design for industrial applications involving soft, yet mechanically robust systems.

**Acknowledgment**

AYa.M would like to acknowledge financial support from the Russian Science Foundation, Grant No 23-69-10001, https://rscf.ru/en/project/23-69-1000. YMJ would like to acknowledge financial support from the Science and Engineering Research Board, Government of India (Grant Nos. CRG/2022/004868 and JCB/2022/000040).